\documentclass[12pt,twoside,a4paper]{article}
\usepackage{qmc,epsfig}

\begin{document}

\title{
Equation of state of strongly coupled\\
QCD at finite density}

\author{Yasuo Umino}
\instit{
Departamento de F\'{\i}sica Te\`orica, Universitat de Val\`encia \\
E--46100 Burjassot, Val\`encia, Spain}
\gdef\theauthor{Yasuo Umino}
\gdef\thetitle{Equation of state of strongly coupled
QCD at finite density}

\maketitle

\begin{abstract}
Using an effective strongly coupled lattice QCD Hamiltonian and Wilson 
fermions we calculate the equation of state for cold and dense quark matter 
by constructing an ansatz which
exactly diagonalizes the Hamiltonian to second order in field operators for
all densities. This ansatz obeys the free lattice Dirac equation with a
chemical potential term and a mass term which is interpreted as the dynamical quark mass. 
We find that the order of chiral phase transition depends on the values of input 
parameters. In the phase with spontaneously broken chiral symmetry the quark Fermi sea 
has negative pressure indicating its mechanical instability. This result is
in qualitative agreement with those obtained using continuum field theory models
with four--point interactions. 
\end{abstract}

\section{Introduction}
\label{intro}

Simulating finite density QCD is one of the outstanding problems in lattice gauge 
theory \cite{cre00}. Because of the sign problem no reliable numerical simulations 
of finite density QCD with three colors exist even in the strong coupling 
limit \cite{alo00}.\footnote{Simulation of two--color four--flavour QCD at finite quark 
number density has recently been reported in \cite{kog01}.} This is a rather 
frustrating situation in view of the current intense interest in finite density QCD 
fueled by the phenomenology of heavy ion collisions, neutron stars, early universe and 
color superconductivity. Therefore even a qualitative description of strongly coupled 
QCD at finite density using field theoretical methods is quite welcome. 

Strongly coupled QCD at finite quark chemical potential $\mu$ has previously been studied 
analytically both in the Euclidean \cite{dam85,ilg85,bil92} and in the Hamiltonian 
\cite{ley88,gre00} formulations. Except for \cite{ley88}, the consensus is that at 
zero or low temperatures strong coupling QCD undergoes a first order chiral phase
transition from the broken symmetry phase below a critical chemical potential $\mu_C$ 
to a chirally symmetric phase above $\mu_C$. In this work we calculate the equation of state
of strongly coupled QCD at finite $\mu$ using Wilson fermions and show that in the phase
with spontaneously broken chiral symmetry the quark Fermi sea has negative pressure 
indicating its mechanical instability. This result is in qualitative agreement with 
those obtained using continuum field theory models with four--point interactions 
\cite{bub96,alf98}.

\section{Effective Hamiltonian and the ansatz for finite $\mu$}
\label{hamiltonian}

We begin by defining our effective Hamiltonian for strongly coupled QCD which was 
first derived by Smit \cite{smi80}. Henceforth we adopt the notation of Smit \cite{smi80}, 
set the lattice spacing to unity ($a = 1$) and work in momentum space. The charge conjugation 
symmetric form of Smit's effective Hamiltonian with a chemical potential $\mu_0$ in momentum 
space is 
\begin{eqnarray}
H_{\rm eff}
& = &
\frac{1}{2} \sum_{\vec{p}}
\biggl[ M_0 \left( \gamma_0\right)_{\sigma\nu} - \mu_0 \delta_{\sigma\nu} \biggr] 
\biggl[ \bigl(\Psi^\dagger_{a\alpha}\bigr)_\sigma(\vec{p}\,),
\bigl(\Psi_{a\alpha}\bigr)_\nu(-\vec{p}\,) \biggr]_-
\nonumber \\
&   & -\frac{K}{8N_{\rm c}} \sum_{\vec{p}_1,\ldots,\vec{p}_4 }\sum_{l=1}^3
\delta_{\vec{p}_1+\cdot\cdot\cdot+\vec{p}_4,\vec{0}}
\Biggl[ e^{i((\vec{p}_1+\vec{p}_2)\cdot \hat{n}_l)} 
+ e^{i((\vec{p}_3+\vec{p}_4)\cdot \hat{n}_l)} \Biggr]
\nonumber \\
&   &
\!\!\!\!\!\!\!\!\!\!\!\!\!\!\!
\otimes \Biggl[
\left( \Sigma_l \right)_{\sigma\nu} 
\bigl(\Psi^\dagger_{a\alpha}\bigr)_\sigma(\vec{p}_1\,)  
\bigl( \Psi_{b\alpha}\bigr)_\nu(\vec{p}_2\,)
-
\left( \Sigma_l \right)^\dagger_{\sigma\nu}
\bigl(\Psi_{a\alpha}\bigr)_\nu(\vec{p}_1\,)
\bigl( \Psi^\dagger_{b\alpha}\bigr)_\sigma(\vec{p}_2\,) \Biggr]
\nonumber \\
&   &
\!\!\!\!\!\!\!\!\!\!\!\!\!\!\!
\otimes \Biggl[
\left( \Sigma_l \right)^\dagger_{\gamma\delta} 
\bigl(\Psi^\dagger_{b\beta}\bigr)_\gamma(\vec{p}_3\,)  
\bigl(\Psi_{a\beta}\bigr)_\delta(\vec{p}_4\,)
-
\left( \Sigma_l \right)_{\gamma\delta}
\bigl(\Psi_{b\beta}\bigr)_\delta(\vec{p}_3\,)
\bigl( \Psi^\dagger_{a\beta}\bigr)_\gamma(\vec{p}_4\,) \Biggr]
\label{eq:heffp}
\end{eqnarray}
where $\left(\Sigma_l \right) = -i \left( \gamma_0 \gamma_l - ir\gamma_0 \right)$
with the Wilson parameter $r$ taking on values between 0 and 1. In the above
Hamiltonian color, flavor and Dirac indices are denoted by $(a, b)$, 
$(\alpha, \beta)$ and $(\sigma, \nu, \gamma, \delta)$, respectively, and summation
convention is implied. Three of the four parameters in this
theory are the Wilson parameter $r$, the current quark mass $M_0$ and the effective
coupling constant $K$ which behaves as $1/g^2$ with $g$ being the QCD coupling
constant. The fourth parameter $\mu_0$ is in general {\em not} the total chemical potential 
$\mu_{\rm tot}$ of the interacting many body system. As shown below the interaction will 
induce a contribution to $\mu_{\rm tot}$ which is momentum dependent. We shall therefore 
refer $\mu_0$ as the "bare" chemical potential. 

Our strategy is to construct an ansatz which obeys the equation of motion corresponding 
to $H_{\rm eff}$ at finite $\mu$ and use it to calculate the equation of state of cold 
and dense quark matter. In free space this ansatz has the same structure as the free lattice 
Dirac field and obeys the free lattice Dirac equation with a mass term which is interpreted 
as the dynamical quark mass \cite{umi00}. Temporarily dropping color and flavor indices, 
this ansatz is given by
\begin{equation}
\Psi_\nu(t,\vec{p}\,) = 
b(\vec{p}\,) \xi_\nu(\vec{p}\,)e^{-i\omega(\vec{p}\,) t}
+ d^\dagger(-\vec{p}\,)\eta_\nu(-\vec{p}\,)e^{+i\omega(\vec{p}\,) t}
\label{eq:psip}
\end{equation}
The annihilation operators for particles, $b$, and anti--particles, $d$, annihilate an
interacting vacuum state and obey the free fermion anti--commutation
relations. The properties of the spinors $\xi$ and $\eta$ are given in \cite{umi00}. 
The equation of motion for a free lattice Dirac field fixes the excitation energy 
$\omega(\vec{p}\,)$ to be
\begin{equation}
\omega(\vec{p}\,) = \left( \sum_l {\rm sin}^2 (\vec{p}\cdot\hat{n}_l)
+ M^2(\vec{p}\,) \right)^{1/2}
\label{eq:EXEN}
\end{equation}
where $M(\vec{p}\,)$ plays the role of the dynamical quark mass or, equivalently, the 
chiral gap.

We now introduce temperature $T$ and $\mu$ simulatneously into our formalism
and take the $T \rightarrow 0$ limit to construct an ansatz for finite $\mu$. 
This is accomplished by subjecting the $b$ and $d$ operators in Eq.~(\ref{eq:psip}) to 
a generalized thermal Bogoliubov transformation as in thermal field dynamics \cite{umezawa}
\begin{eqnarray}
b(\vec{p}\:) 
& = & 
\alpha_p B(\vec{p}\:) - \beta_p \tilde{B}^{\dagger}(-\vec{p}\:)
\label{eq:BTb} \\
d(\vec{p}\:) 
& = &
\gamma_p D(\vec{p}\:) - \delta_p \tilde{D}^{\dagger}(-\vec{p}\:)
\label{eq:TBT}
\end{eqnarray}
The thermal field operators $B$ and $\tilde{B}^\dagger$
annihilate a quasi--particle and  create a quasi--hole at finite $T$ and
$\mu$, respectively, while $D$ and  $\tilde{D}^\dagger$ are the
annihilation operator for a quasi--anti--particle  and creation opertor for
a quasi--anti--hole, respectively. 

The thermal annihilation operators annihilate the interacting thermal vacuum 
state $|\,{\cal G}(T,\mu)\rangle$ {\em for each $T$ and $\mu$}.  
\begin{eqnarray}
B(\vec{p}\:) |\,{\cal G}(T,\mu)\rangle 
=\tilde{B}(\vec{p}\:) |\,{\cal G}(T,\mu)\rangle 
& = & 
0 \label{eq:AVAC1} \\
D(\vec{p}\:) |\,{\cal G}(T,\mu)\rangle 
=\tilde{D}(\vec{p}\:) |\,{\cal G}(T,\mu)\rangle 
& = & 
0 \label{eq:AVAC2}
\end{eqnarray}
The thermal doubling of the Hilbert space accompanying the thermal
Bogoliubov transformation is implicit in Eqs.~(\ref{eq:AVAC1}) and (\ref{eq:AVAC2})
where a vacuum state which is annihilated by thermal operators 
$B$, $\tilde{B}$, $D$ and $\tilde{D}$ is defined. Since we
shall be working only in the space of quantum field operators it 
is not necessary to specify the structure of $|\,{\cal G}(T,\mu)\rangle$. 

The thermal operators satisfy the Fermion anti--commutation relations 
just as the $b$ and $d$ operators in the free space ansatz. The coefficients of the 
transformation are $\alpha_p = \sqrt{1-n_p^-}$,
$\beta_p = \sqrt{n_p^-}$, $\gamma_p = \sqrt{1-n_p^+}$ and
$\delta_p = \sqrt{n_p^+}$, where 
$n_p^{\pm}= [e^{(\omega_p \pm \mu)/(k_B T)}+1]^{-1}$ are the Fermi 
distribution functions for particles and anti--particles.
They are chosen so that the total particle number densities are given by
\begin{eqnarray}
n_p^- 
& = &
\langle \,{\cal G}(T,\mu)|b^{\dagger}(\vec{p}\:)b(\vec{p}\:)
|\,{\cal G}(T,\mu)\rangle\\
n_p^+
& = &
\langle \,{\cal G}(T,\mu)|d^{\dagger}(\vec{p}\:)d(\vec{p}\:)
|\,{\cal G}(T,\mu)\rangle %
\label{eq:DIST}
\end{eqnarray}
Hence in this approach temperature and chemical
potential are introduced simultaneously through the coefficients of the
thermal Bogoliubov transformation and are treated on an equal
footing. We stress that the chemical potential appearing in the
Fermi distribution functions is the {\em total} chemcial potential of the
interacting many body system which is not necessarily
equal to the bare chemical potential $\mu_0$ appearing in the effective
Hamiltonian in Eq.~(\ref{eq:heffp}).

We demand that our ansatz satisfies the equation of motion corresponding to the free 
lattice Dirac Hamiltonian with a chemical potential term. This Hamiltonian is given by
\begin{eqnarray}
H^0 
& = & 
\frac{1}{2} \sum_{\vec{p}}
\Biggl[ - \sum_l {\rm sin}(\vec{p}\cdot\hat{n}_l)(\gamma_0\gamma_l)_{\sigma\nu}
+ M(\vec{p}\,)(\gamma_0)_{\sigma\nu} - \mu_{\rm tot} \delta_{\sigma\nu} \Biggr]
\nonumber \\
&   &
\;\;\;\;\;\;
\otimes\Bigg[ \Psi_\sigma^\dagger(t, \vec{p}\,), \Psi_\nu(t, \vec{p}\,) \Biggr]_-
\label{eq:H0}
\end{eqnarray}
As in \cite{umi00} the mass $M(\vec{p}\,)$ will be identified with the chiral gap while
$\mu_{\rm tot}$ is the total chemical potential mentioned above. In the $T \rightarrow 0$ 
limit the excitations of anti--holes are supressed which amount to setting $\gamma_p = 1$ 
and $\delta_p = 0$ in Eq.~(\ref{eq:TBT}). Thus our ansatz at finite $\mu_{\rm tot}$ is given 
by
\begin{eqnarray}
\Psi_\nu(t, \vec{p}\:) 
& = &
\biggl[ \alpha_p B(\vec{p}\:) - \beta_p \tilde{B}^{\dagger}(-\vec{p}\:)\biggr]
\xi_\nu(\vec{p}\,)e^{-i[\omega(\vec{p}\,)-\mu_{\rm tot}] t}  \nonumber \\
&    &
\;\;\;\;\;\;\;\;\;\;\;\;\;\;\;\;\;\;\;\; +\;\;
\gamma_p D^{\dagger}(-\vec{p}\:) \eta_\nu(-\vec{p}\,)e^{+i[\omega(\vec{p}\,) +
\mu_{\rm tot}] t}
\label{eq:HAAGTMU}
\end{eqnarray}
The spinors $\xi$ and $\eta$ obey the same properties as
in free space and the excitation energy $\omega(\vec{p}\,)$ has the same
form as in Eq.~(\ref{eq:EXEN}). 

Two quantities of interest that can be calculated in a straightforward manner
using Eq.~(\ref{eq:HAAGTMU}) are the quark number density 
$\langle \bar{\Psi}\gamma_0 \Psi \rangle$ and the chiral condensate 
$\langle \bar{\Psi} \Psi \rangle$. They are found to be 
\begin{eqnarray}
\langle \bar{\Psi}\gamma_0 \Psi \rangle 
& = &
\sum_{\vec{p}} \beta_p^2 
\label{eq:qnum}
\end{eqnarray}
and
\begin{eqnarray}
\langle \bar{\Psi}\Psi \rangle 
& = & 
-\sum_{\vec{p}} \alpha_p^2 \frac{M(\vec{p}\,)}{\omega(\vec{p}\,)}
\label{eq:cond}
\end{eqnarray}
We see immediately from Eq.~(\ref{eq:cond}) that 
the chiral condensate is proportional to the dynamical quark mass and therefore 
can clearly be identified as being the order parameter for the chiral phase 
transition at finite density.

\section{Application of the Equation of Motion}
\label{diagonal}
\subsection{The Gap Equation and the Induced Chemical Potential}
\label{gapeq}

We proceed by using the ansatz shown in Eq.~(\ref{eq:HAAGTMU}) in our effective 
Hamiltonian to determine the equation of motion. This is accomplished 
by exploiting the fact that our ansatz satisfies the equation of motion corresponding to 
the free Hamiltonian at finite $\mu$ given in Eq.~(\ref{eq:H0}). We therefore have 
the relation
\begin{equation}
:\Bigl[ \bigl(\Psi_{a\alpha}\bigr)_\nu(t, \vec{q}\,), H^0 \Bigr]_- :
\;\;\;\;
=
\;\;\;\;
:\Bigl[ \bigl(\Psi_{a\alpha}\bigr)_\nu(t, \vec{q}\,), H_{\rm eff} \Bigr]_- :
\label{eq:CRUX}
\end{equation}
where the symbol :  : denotes normal ordering with respect to the thermal vacuum 
$|\,{\cal G}(T=0,\mu)\rangle$. Evaluating both sides of Eq.~(\ref{eq:CRUX}) and equating 
terms which are linear in the field operators we obtain
\begin{eqnarray}
\lefteqn{ \!\!\!\!\!\!\!\!\!\!\!\!\!\!\!
\Bigl[ \sum_l {\rm sin}(\vec{q}\cdot\hat{n}_l)(\gamma_0\gamma_l)_{\nu\delta}
+ M(\vec{q}\,)(\gamma_0)_{\nu\delta} - \mu_{\rm tot}\delta_{\nu\delta} \Bigr]
\bigl(\Psi_{a\alpha}\bigr)_\delta(t, \vec{q}\,) =
} \nonumber \\
&   &
\Bigl[ A(\vec{q}\,)(\gamma_0\gamma_l)_{\nu\delta} +
B(\vec{q}\,)(\gamma_0)_{\nu\delta} + C(\vec{q}\,)\delta_{\nu\delta} \Bigr]
\bigl(\Psi_{a\alpha}\bigr)_\delta(t, \vec{q}\,)
\label{eq:EOM2}
\end{eqnarray}
The momentum dependent coefficients $B$ and $C$ in Eq.~(\ref{eq:EOM2}) determine the 
gap equation and the total chemical potential, respectively.
\begin{eqnarray}
M(\vec{q}\,)
& = &
B(\vec{q}\,)
\nonumber \\
& = &
M_0 + \frac{3}{2}K(1 - r^2) \sum_{\vec{p}} \left( 1 - \beta_p^2 \right)
\frac{M(\vec{p}\,)}{\omega(\vec{p}\,)}
\nonumber \\
&   &
\!\!\!\!\!\!\!\!\!\!
+ \;\;\frac{1}{N_{\rm c}} K\sum_{\vec{p},l} \left( 1 - \beta_p^2 \right)
\frac{M(\vec{p}\,)}{\omega(\vec{p}\,)}
\Biggl\{
8r^2  \cos(\vec{p}\cdot\hat{n}_l) \cos(\vec{q}\cdot\hat{n}_l)
\nonumber \\
&   &
\!\!\!\!\!\!\!\!\!\!
- \; \frac{1}{2}(1+r^2) \cos(\vec{p}+\vec{q}\, )\cdot\hat{n}_l
\Biggr\}
\label{eq:GAPEQ}
\end{eqnarray}
\begin{eqnarray}
\mu_{\rm tot}(\vec{q}\,)
& = &
- C(\vec{q}\,) \nonumber \\
& = &
\!\!\mu_0\! +\! \frac{1}{4} \frac{K}{N_c}\! \sum_{\vec{p},l} \beta_p^2
\Bigl[ 2N_c \left( 1+r^2 \right) - 2 \left( 1-r^2 \right)
{\rm cos}\left( \vec{p} + \vec{q}\, \right) \Bigr]
\label{eq:MUTOT}
\end{eqnarray} 

The structure of the gap equation Eq.~(\ref{eq:GAPEQ}) is very similar to that
in free space ($\beta^2_p = 0$) found in \cite{umi00}. The dynamical quark
mass is a constant to lowest order in $N_c$ but becomes momentum dependent
once $1/N_c$ correction is taken into account. We see from Eq.~(\ref{eq:MUTOT}) 
that the total chemical potential is a sum of the bare chemical potential $\mu_0$ 
and a momentum dependent contribution generated by the interaction. Because of this 
induced chemical potential Eqs.~(\ref{eq:GAPEQ}) and (\ref{eq:MUTOT}) are coupled 
and must be solved self--consistently. It should be noted that this shifting of the 
bare chemical potential by the interaction is not a new effect. In the 
Nambu--Jona--Lasinio model \cite{nam61} within the Hartree--Fock approximation and 
at finite $T$ and $\mu$, the interaction induces a contribution to the total chemcial 
potential which is proportional to the number density \cite{asa89}. 

\begin{figure}[tbp]
\begin{center}
\epsfig{file=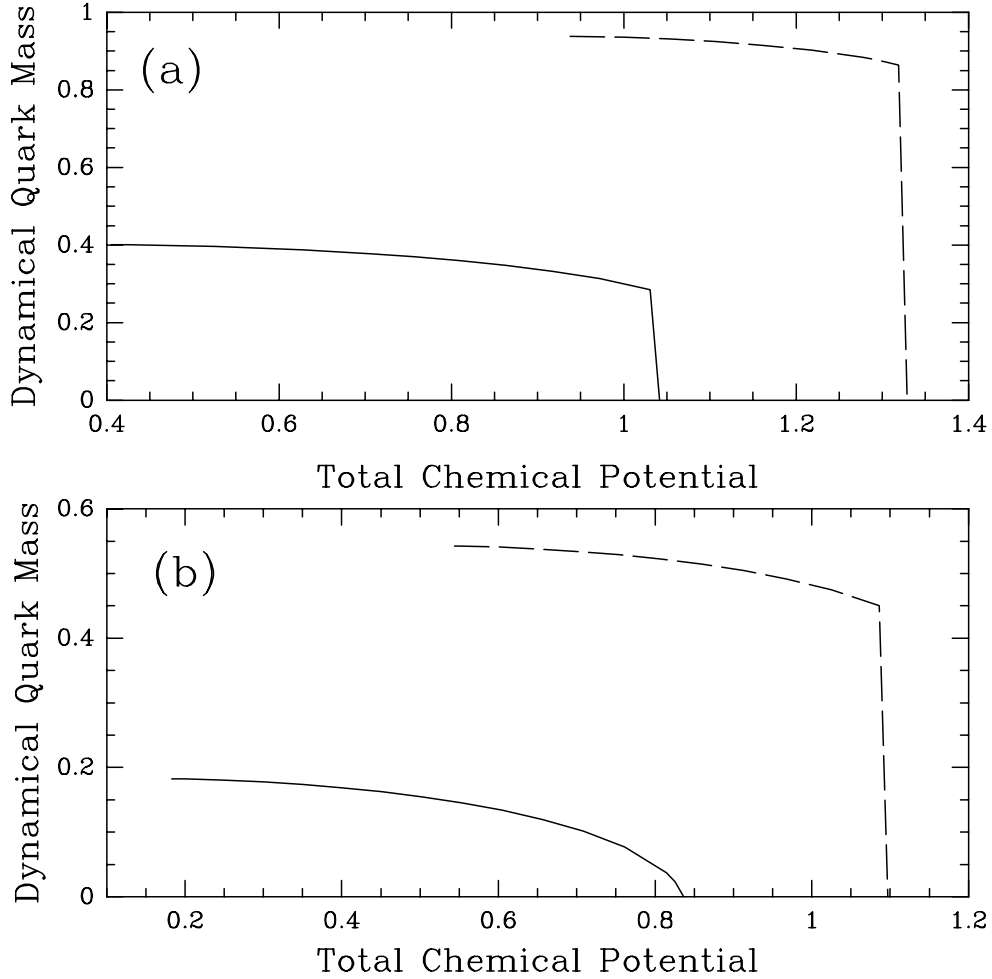,width=13cm,height=13cm}
\end{center}
\caption{a) Dynamical quark mass versus the total chemical potential to lowest order in 
$N_c$ for $K$ = 0.8 (solid line) and 1.0 (dashed line) with $r$ = 0. A first order chiral 
phase transition takes places in both cases with a critical chemical potential of 
$\mu_C$ = 1.042 and 1.329 for $K$ = 0.8 and 1.0, respectively. 
b) Dynamical quark mass versus the total chemical potential to lowest order in $N_c$ for 
$K$ = 0.8 (solid line) and 0.9 (dashed line) with $r$ = 0.25. 
A second order chiral phase transition takes place for $K$ = 0.8 with a critical
chemical potential of $\mu_C$ = 0.836 while the transition is first order for 
$K$ = 0.9 with $\mu_C$ = 1.086. The current quark mass is set to zero 
($M_0 =0$) in all cases.} 
\label{fig1}
\end{figure}

Solutions for the coupled equations Eqs.~(\ref{eq:GAPEQ}) and (\ref{eq:MUTOT}) to 
lowest order in $N_C$ are shown in Figure~\ref{fig1} where we present the dynamical quark 
masses as functions of the total chemical potential for various combinations of the Wilson 
parameter $r$ and the coupling constant $K$ with a vanishing current quark mass. 
As shown in Figure~1a we find first order chiral phase transitions when $r$ = 0 
({\em i.e.} with naive fermions) with critical chemical potentials of $\mu_C \approx 1.04$ 
and 1.33 for $K$ = 0.8 and 0.9, respectively. As far as the order of the phase transition 
is concerned, our results agree with those obtained using 
the Euclidean formulation with Kogut--Susskind fermions \cite{dam85,ilg85,bil92} and also 
with a recent result obtained with naive fermions ($r$ = 0) in a Hamiltonian formulation 
\cite{gre00}. However we disagree with the result of Le~Yaouanc et al. \cite{ley88} who find 
a second order chiral phase transition using the {\em same} effective Hamiltonian as in this
work and  naive fermions.

For $r$ = 0.25 we see from Figure~1b that the phase transition can be either first or 
second order depending on the value of the effective coupling constant $K$. When $K$ = 0.8 
we find a second order phase transition with $\mu_C \approx 0.84$ while if the coupling 
constant is increased to 0.9 the phase transition becomes first order with a
larger critical chemical potential of $\mu_C \approx 1.09$. It would be interesting to
investigate whether the approaches adapted in \cite{ley88} and \cite{gre00} to study 
strongly coupled QCD in the Hamiltonian lattice formulation using naive fermions would  
result in a similar behaviour of the order of the phase transition when extended to 
$r > 0$. Finally, we note that for both $r$ = 0 and 0.25 the value of the critical chemical 
potential increases as $K$ is increased. The same behaviour has been observed in \cite{bil92}.

\subsection{Off--diagonal Hamiltonian and the Vacuum Energy Density}
\label{offham}

The off--diagonal Hamiltonian which is bilinear in field creation and annihilation 
operators may be written as
\begin{equation}
H_{\rm off}|\,{\cal G}(0, \mu)\rangle = 
\sum_{\vec{q}} F(\vec{q}\,) B^\dagger_{\alpha,a}(\vec{q}\,)D^\dagger_{\alpha,a}(-\vec{q}\,)
|\,{\cal G}(0, \mu)\rangle
\label{eq:hoff2}
\end{equation}
where $F(\vec{q}\,)$ is a scalar function of $\vec{q}$. We see from Eq.~(\ref{eq:hoff2}) that 
the elementary excitation of the effective Hamiltonian involves color singlet (quasi--) 
quark--anti--quark excitations coupled to zero total three momentum. With the use of the
equation of motion and properties of $\eta$ spinor \cite{umi00} it can be shown analytically 
that $H_{\rm off}|\,{\cal G}(0, \mu)\rangle = 0$ for all $\mu$. Therefore our ansatz exactly 
diagonalizes the effective Hamiltonian to second order in field operators for all densities.

Having verified that our ansatz diagonalizes the second order Hamiltonian we are in a position 
to evaluate the vacuum energy density. Using Eq.~(\ref{eq:EOM2}) we find
\begin{eqnarray}
\lefteqn{
\frac{1}{V} \langle {\cal G}(0, \mu)| H_{\rm eff} |\,{\cal G}(0, \mu) \rangle =} \nonumber \\
&  &
-2 N_c \sum_{\vec{p}} \Biggl\{
\alpha_p^2 \Biggl[ \frac{3}{2} K (1 + r^2) + \omega(\vec{p}\,)
+ \frac{M(\vec{p}\,)}{\omega(\vec{p}\,)} M_0
\nonumber\\
&  &
-\frac{1}{N_c} \frac{K}{2} (1 - r^2) \sum_{\vec{q}}
\cos(\vec{p} + \vec{q}\,)\cdot\hat{n}_l - \mu_{\rm tot} \Biggr]
+ (1 + \beta_p^2) \mu_0 \Biggr\}
\label{eq:VED}
\end{eqnarray}
From Eq.~(\ref{eq:VED}) we can evaluate the difference of the vacuum energy 
densities in the Wigner--Weyl ($M(\vec{q}\,) = 0$) and Nambu--Goldstone 
($M(\vec{q}\,) \neq 0$) phases of the theory. We find
\begin{equation}
\Delta E = 
\frac{1}{V}\langle {\cal G}|H_{\rm eff}|\,{\cal G} \rangle |_{M(\vec{q}\,) = 0}
- \frac{1}{V}\langle {\cal G}|H_{\rm eff}|\,{\cal G} \rangle |_{M(\vec{q}\,) \neq 0}
> 0
\end{equation}
and therefore the true ground state of our interacting many body system is in the
phase with broken chiral symmetry.

\section{Equation of State}
\label{eos}

\begin{figure}[tbh]
\begin{center}
\epsfig{file=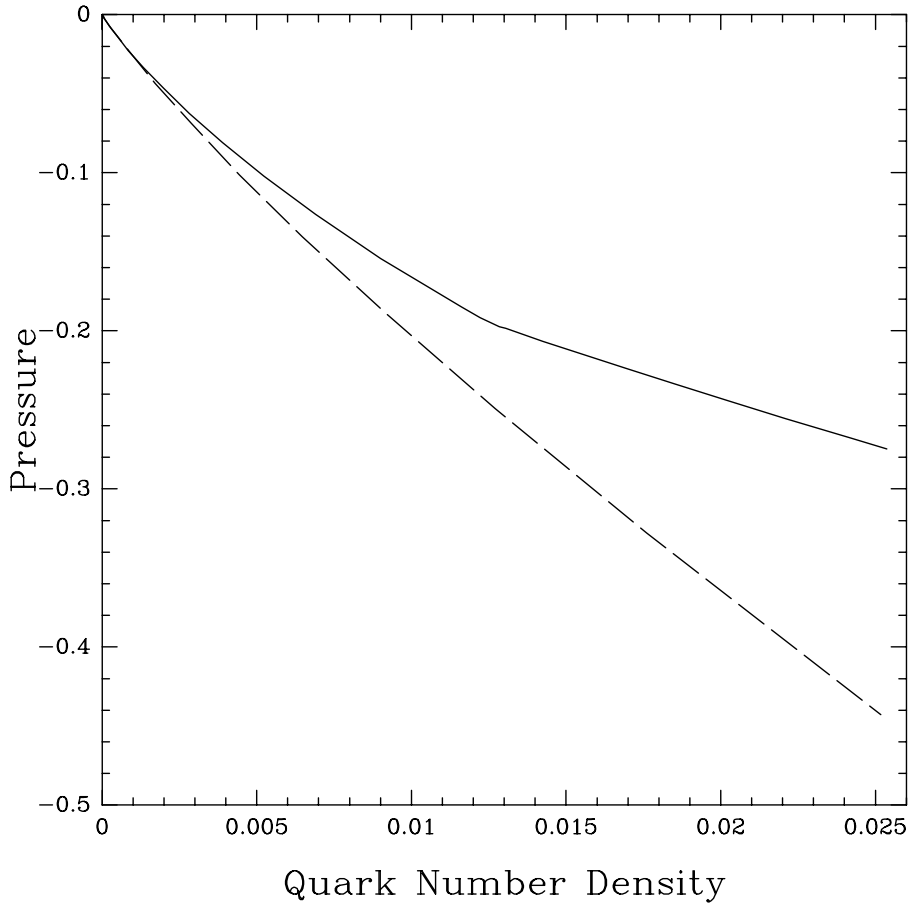,width=12cm,height=12cm}
\end{center}
\caption{Pressure versus the quark number density to lowest order in $N_c$ 
for $K$ = 0.8 (solid line) and 0.9 (dashed line) with $M_0 = 0$ and $r$ = 0.25. 
Quark number densities at the chiral phase transtion point are $n_q$ = 0.013 
and 0.025 for $K$ = 0.8 and 0.9, respectively.} 
\label{fig2}
\end{figure}

Having solved the gap equation for the dynamical quark mass we proceed to 
determine the equation of state of cold and dense quark matter. We present results 
obtained with a vanishing current quark mass and Wilson fermions for $r$ = 0.25 
to leading order in $N_c$ since in this case we observe both first and second order 
chiral phase transitions. The pressure of the cold and dense quark matter is obtained 
by numerically evaluating the thermodynamic potential density. In Figure~\ref{fig2} 
we plot the pressure as functions of quark number density for $K$ = 0.8 and 0.9. In both 
cases we find the pressure to be negative in the broken symmetry phase and therefore 
the quark Fermi sea is mechanically unstable with an imaginary speed of sound.

Our conclusion regarding the quark matter stability at finite density is consistent with 
results obtained using the Nambu--Jona--Lasinio model \cite{bub96} and the effective
instanton induced 't~Hooft interaction model \cite{alf98}. In both \cite{bub96} and 
\cite{alf98} mean field calculations show that cold and dense quark matter is unstable 
in the phase with spontaneously broken chiral symmetry and lead the authors to speculate
the formation of quark droplets reminiscent of the MIT bag model.

\section{Conclusion}
\label{Section6}

In this work we studied strongly coupled QCD in the Hamiltonian lattice formalism at finite 
density using Wilson fermions. Starting from an effective
Hamiltonian we constructed and an ansatz which exactly diagonalizes the Hamiltonian to
second order in field operators for all densities. This ansatz obeys the free lattice Dirac 
equation with a chemical potential term and a mass term which plays the role of the dynamical 
quark mass. This mass and the total chemical potential of the interacting many body system 
were determined to lowest order in $N_c$ by solving a coupled set of equations 
obtained from the equation of motion. We find a first order chiral phase transition 
with naive fermions while the order of the phase transition can be either first or 
second for Wilson fermions. As in earlier studies using continuum four--fermion
interaction models we find that the quark Fermi sea is mechanically unstable in the phase 
with broken chiral symmetry. Our results for the equation of state of cold and dense quark 
matter should certainly be verified in any future numerical simulation of finite density QCD.


\end{document}